\documentclass[runningheads]{llncs}
\usepackage{amsmath}
\DeclareMathOperator*{\argmax}{arg\,max}
\usepackage{graphicx}
\usepackage{commath}
\usepackage{cite}
\begin{document}
\title{Distributed Vector Representations of Folksong Motifs}
%
%
\author{Aitor Arronte Alvarez\inst{1} \and
Francisco G\'omez-Martin\inst{2}}
\authorrunning{A. Arronte-Alvarez \and F. G\'omez-Martin}
%
\institute{ 
Center for Language and Technology, University of Hawaii at Manoa, United States\\
\email{arronte@hawaii.edu}\\
\and
Applied Mathematics Department, Technical University of Madrid, Spain\\
\email{fmartin@etsisi.upm.es}}
\maketitle              
\begin{abstract}
This article presents a distributed vector representation model for learning folksong motifs. A skip-gram version of word2vec with negative sampling is used to represent high quality embeddings. Motifs from the Essen Folksong collection are compared based on their cosine similarity. A new evaluation method for testing the quality of the embeddings based on a melodic similarity task is presented to show how the vector space can represent complex contextual features, and how it can be utilized for the study of folksong variation.

\end{abstract}
\section{Introduction}

Vector representations of words have been widely used in Natural Language Processing (NLP) tasks \cite{rumelhart1986learning}. Following the distributional hypothesis \cite{harris1954distributional}, \cite{lappin2015handbook}, vector space models represent, or embed, words that are semantically related to each other closer in a continuous vector space \cite{turney2010frequency}. A recent development in vector space models is word2vec    
\cite{mikolov2013efficient,mikolov2013distributed, goldberg2014word2vec}, developed  for learning high-quality word vectors from large corpora. 

A neural network language model for learning word-embeddings was first proposed to learn a statistical language model and a word vector representation \cite{bengio2003neural}. A simpler model using a neural net with a single hidden layer to learn word vector representations, and then train a language model was later developed \cite{mikolov2009neural}. Word2vec follows this simpler approach in two steps: first, continuous word vectors are learned using the simpler model \cite{mikolov2009neural}, and then an n-gram is trained using these representations.

The relation between music and language has been studied in the cognitive science literature. Even though they are treated as different cognitive faculties, both share structural characteristics and generate similar expectations on the listener \cite{besson2001comparison}. NLP methods have been adapted and adopted in Music Information Retrieval (MIR) contexts \cite{conklin1995multiple}, \cite{boulanger2012modeling}, \cite{Boom2017LargescaleUM}. More specifically, word2vec was used to model musical contexts in western classical music works \cite{herremans2017modeling}, and for chord recommendations \cite{huang2016chordripple}. In both cases the music compositions studied were complex polyphonic works. The work presented in this article uses a much less data intensive material: monophonic songs. 

Following the distributional hypothesis in semantics, the goal of this research is to adopt the skip-gram version of the word2vec model for the distributional representation of melodic units. Several melodic features such as contour, grouping, and small size motifs seem to be part of the so called ‘Statistical Music Universals’ \cite{nettl2000ethnomusicologist}, \cite{savage2015statistical}. This sequential processing of melodic units may be related to the human capacity to group and comprehend motifs as units within a melodic context. Our hypothesis is that these units may relate to each other in a melody in similar ways as words do in sentences. If that is the case, the distributional hypothesis should hold true for folksong melodies.

In the following sections a description of the skip-gram version of word2vec to learn motifs from the Essen Folksong Collection \cite{schaffrath1995essen} is presented. We will present different similarity measures to determine how melodic context can capture the similarity of folksong motifs.

\section{Word2vec: Representing Folksong Motifs in a Distributed Vector Space }
\subsection{Word2vec Model}
In the skip-gram version of the word2vec model, the goal is to find word embeddings that can predict the surrounding words of a target word in a sentence or document \cite{mikolov2013distributed}. Formally, the model can be defined in the following terms: given a corpus $W$ of words $w$ and contexts $c$, the network tries to predict the surrounding words of a target in a context. The objective of the skip-gram is to maximize the following log probability:
\begin{equation}
\argmax_\theta \prod_{w  \in W} \Bigg[  \prod_{c  \in C}  p(c \mid w; \theta) \Bigg]
\end{equation}
where  $p(c \mid w; \theta)$ is calculated by the softmax function:

\begin{equation}
p(c \mid w; \theta)= \frac{e^{v_c \cdot v_w}} {\sum\limits_{c\sp{\prime} \in C} e^{v_{c\sp{\prime} } \cdot v_w}}
\end{equation}
where \(v_c\) and \( v_w \in R^d \) are vector representations of \textit{v} and \textit{c}, and \textit{C} is the set of all possible contexts. The set of parameters \(\theta\) is composed of \( v_{c_i} \), \( v_{w_i} \) for \( w\in W\).

Since the term \( p(w; \theta) \) involves a summation over all possible contexts \( c \sp{\prime} \) becomes computationally very intensive, and it is normally replaced with negative sampling \cite{mikolov2013distributed}. This article uses this sampling technique.

The cosine similarity measure is used to determine the relatedness of two embeddings. The metric for a pair of words \( w_1 \) and \( w_2 \) can be defined as \cite{schnabel2015evaluation} :

\begin{equation}
cos(w_1, w_2)=\frac{\overrightarrow{w_1} \cdot \overrightarrow{w_2}}{\norm{\overrightarrow{w_1}} \norm{\overrightarrow{w_2} }} 
\end{equation}
for all similarity computations in the embedding space, where $\overrightarrow{w}$ is a real-valued vector embedding of word $w$.

\subsection{Melodic context and motif representation}
\label{ssec:melcontext}

We are interested in studying how word2vec can model melodic context using small musical motifs instead of words. In the present research context is understood as the sequential organization of melodic units that establish statistically relevant relationships with one another in a melodic segment.

Melodic similarity and classification methods depend strongly on melodic representation \cite{toiviainen2002computational}. Motifs from the Essen folksong collection are represented by using strings. First, intervals are codified for each song by using Music21 \cite{Cuthbert2010Music21AT} chromatic step values from the original Kern format, and encode interval direction with Boolean values (\texttt{1} for ascending and \texttt{0} for descending). For instance, the string \texttt{21} represents an ascending major second, and the string \texttt{30} a descending minor third. Repeated notes are encoded as \texttt{00}. 

Once the entire folksong corpus is encoded using this scheme, motifs are extracted as multi-words \cite{mikolov2013distributed}. A multi-word is then a concatenation of two or more intervals or durations that are found in a melody adjacent to each other. For example, an intervallic multi-word of size 3 \texttt{30\_00\_21} represents a descending minor third, followed by a repeated note, and by an ascending major second.

The multi-word representation of motifs is obtained following these steps:
\begin{itemize}
\item From a corpus of intervals we create a vocabulary of multi-word    $M$ with multi-words $mw$ of length 2. Only those $mw$ that occur at least 10 times are kept, based on the quality of the results from ad-hoc queries.
\item For each $mw$ in $M$ intervals in the corpus are substituted with their corresponding $mw$.
\end{itemize}

The same procedure is used for \textit{mw} of size 3, with the only difference that the minimum number of occurrences of \textit{mw} in a corpus is set to 5. The word2vec model is run based on the corpora created obtaining vector representations for all the motifs.

\subsection{Evaluation methods} 
\label{ssec:eval}

Evaluation of Word Embeddings (WE) falls into two categories: intrinsic and extrinsic evaluation \cite{schnabel2015evaluation}. Intrinsic evaluation methods test for syntactic or semantic relationships between words using predefined queries. Then, methods are evaluated by aggregating correlation scores. Extrinsic evaluations are performed by using WE as the input feature for another task, and then embeddings are evaluated based on the changes in the performance of that particular task.

This study concentrates on intrinsic evaluations, more specific on relatedness and analogy. Relatedness in WE is the cosine similarity between two words. Pairs of words should have higher correlation scores when compared with human annotated semantic similarity scores \cite{schnabel2015evaluation}. Analogical reasoning was first used for testing semantic relationships between pairs of words given specific phrases: given a term \textit{x}  and a term \textit{y} so that \textit{x:y} resembles a sample relationship \textit{i:j}\cite{mikolov2013efficient}. All these evaluation methods are language specific, and have not being adapted for MIR tasks.

Given the non-linguistic nature of music, and the difficulty of interpreting WE, more so when they represent melodic motifs, a new method is presented for evaluating Melodic Embeddings (ME) based on variations of motifs and similarity measures for those motifs in relation to a reference one. The method proceeds as follows:

\begin{enumerate}
  \item For each multi-word \textit{\( mw_i \)}, where \textit{ i = 1, 2, ..., l} and \textit{l} is the cardinality of the vocabulary \textit{M} from corpus \textit{C}, we compute \textit{max(cos(\( mw_i \), \( mw_j \))) for all j}, and obtain the most related multi-word \textit{\( mw_i^+\)} of \textit{\( mw_i \)}, so that \textit{\( mw_i \)} : \textit{\( mw_i^+\)}, and an unrelated multi-word \textit{\( mw_i^-\)}, where \textit{cos(\( mw_i \), \( mw_i^-\))}\textless \textit{h}, where \textit{h} is an acceptable similarity threshold.
  \item Chose from \textit{C} a melodic segment \textit{c} and replace \textit{\( mw_i\)} with \textit{\( mw_i^+\)} and \textit{\( mw_i^-\)}, obtaining a related \textit{\( c^+\)} and an unrelated \textit{\( c^-\)}  melodic segments. This action is performed for all segments in \textit{C}.
  \item Obtain \textit{sim(c, \( c^+\))} and \textit{sim(c, \( c^-\))}, where \textit{sim()} is a function that computes a measure of melodic similarity between pairs of melodic segments.
\end{enumerate}

The idea behind this evaluation method is that, if vector representations of motifs are of good quality, when a motif \textit{\( mw_i\)} is replaced with its most similar motif \textit{\( mw_i^+\)} in a melodic segment \textit{c} obtaining \textit{\( c^+\)}, then a melodic similarity measure should indicate that segment \textit{c} is more similar to \textit{\( c^+\)} than to \textit{\( c^-\)}. 

To measure intervallic similarity, sequences are evaluated using the mean absolute difference in intervals (\textit{diffint}) \cite{mullensiefen2004cognitive}. Since this study deals with equal-length sequences, note sequences are evaluated with city block distance (\textit{citydist}) \cite{scherrer1971experiment}, and for duration-weighted pitch sequences correlation distance (\textit{corrdist}) \cite{janssen2017finding}. In order to compute distance measures based on note sequences, a vector of pitches represented as numerical MIDI values is used.

\subsection{Evaluating motif embeddings}
\label{ssec:motemb}

A sample of 2000 melodic segments is randomly selected from the European subcollection from the Essen folksong corpus. Multi-word embedings of size 2 and 3 are obtained using the skip-gram version of word2vec with context size of 5 and vector dimension of 150. We measure melodic similarity using \textit{diffint}, \textit{citydist}, and \textit{corrdist} for related and unrelated multi-word melodic segments using the method presented in \ref{ssec:eval}, and compare their means. 

\begin{figure}
  \centering
  \includegraphics[width=1\textwidth]{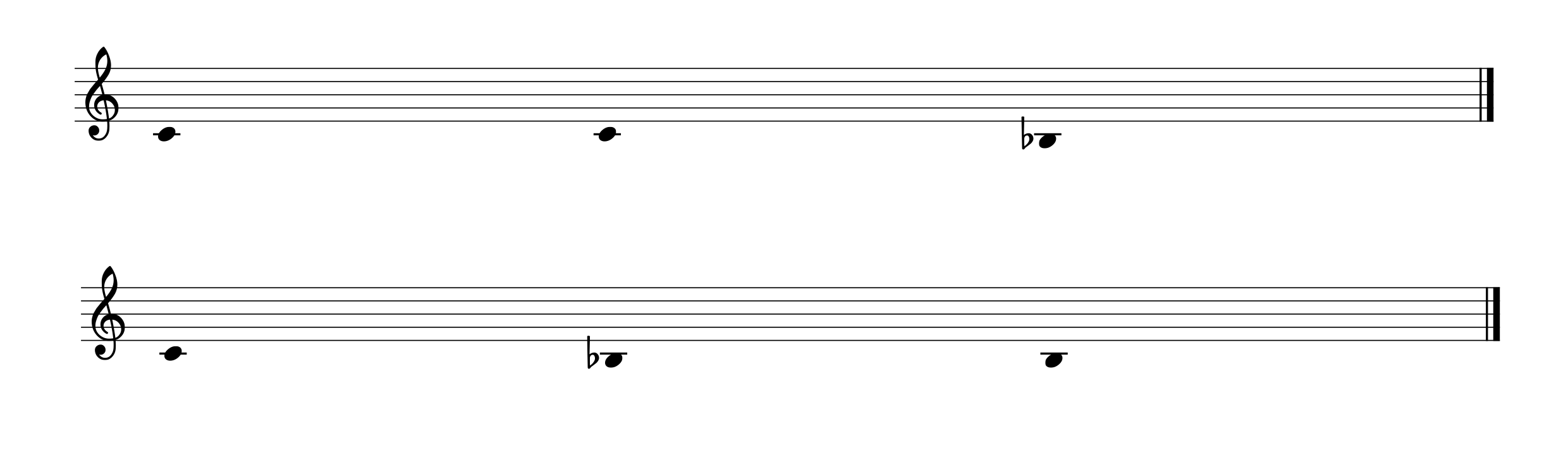}
  \caption{\label{img:motif2}Similar intervallic motifs from \textit{mw} of size 2}
\end{figure}

\begin{figure}
  \centering
  \includegraphics[width=1\textwidth]{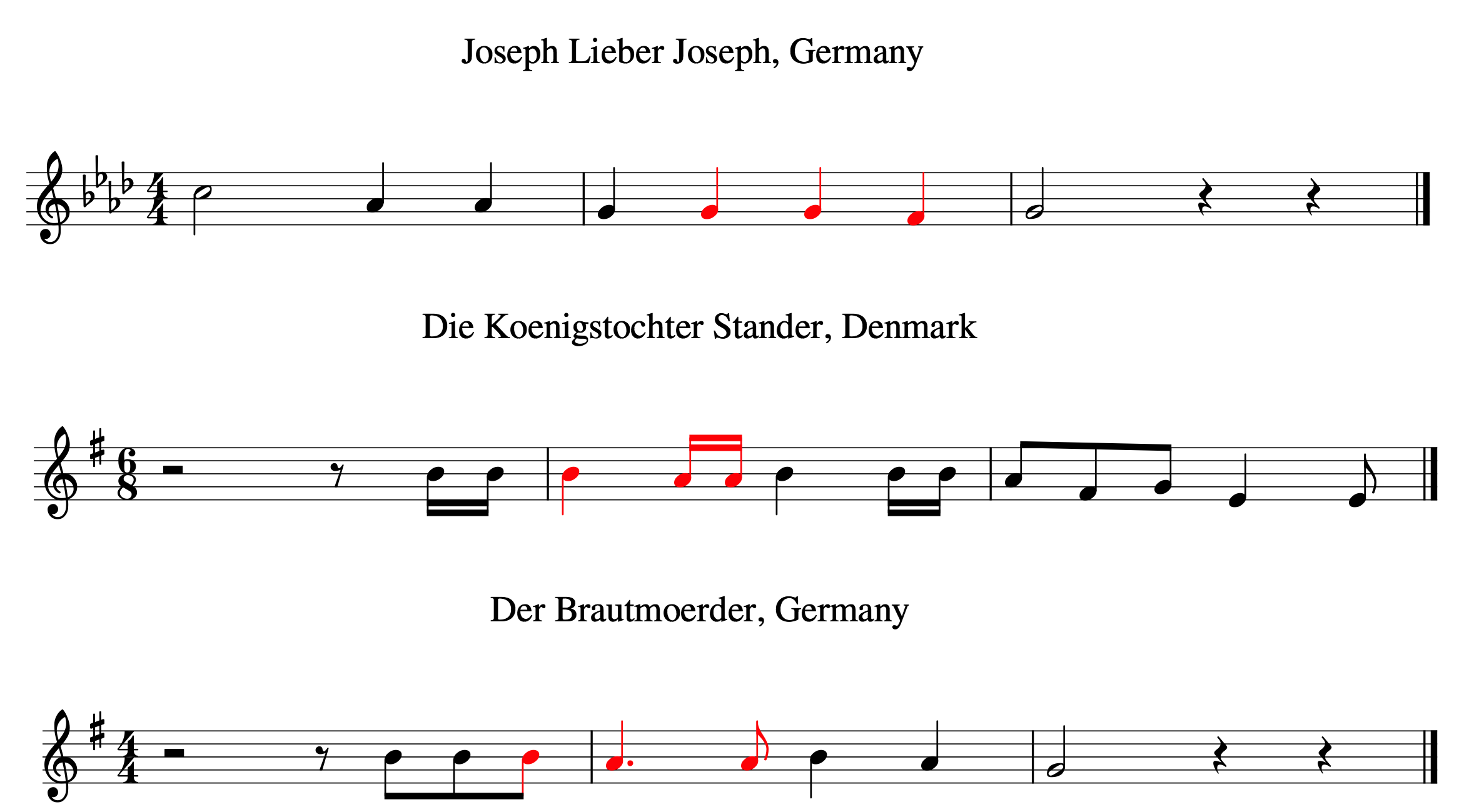}
  \caption{\label{img:motif3}Fragments of European folksongs with similar intervallic motifs colored in red}
\end{figure}

Wilcoxon rank sum test is performed on related and unrelated melodic segments for all similarity measures, resulting on significant differences in means for all measures (\textit{p-value}\textless 0.01). Ad-hoc queries of intervallic motif embeddings of size 2 show similarity between motifs based on the context. For instance, Figure \ref{img:motif2} shows similar motifs from $mw$ of size 2 (transposed to C), and Figure \ref{img:motif3}, shows melodic examples where  those motifs are present in similar melodic contexts: all three fragments contain the target motif, either \texttt{00\_20} or \texttt{20\_00} preceded by a melodic unison and followed by an ascending major second.

Next, closely related and unrelated melodic segments variations from a reference segment using the procedure described in \ref{ssec:eval} are computed. We compare the similarity between a reference melodic segment with its most related variation and the same reference segment with a close variation, and with a non related (or distant) variation. The cosine similarity for multi-words of size 2 and 3 is used to select closely related and unrelated motifs. We utilize the Euclidean distance for comparing the average similarity scores of the 2000 segments and all the variants described.

The results in Table \ref{tab:eudist} show that the distance of the similarity scores between the reference segments and their variations, and the reference segments and closely related variants (\textit{ref\_var\_ref\_close\_var}) yield better results than when we compare the reference segments and  their variants, with the reference segments with distantly related variants (\textit{ref\_var\_ref\_distant\_var}).

\begin{table}
\centering
\begin{tabular}{ |p{2cm}|p{3cm}|p{3cm}|p{2cm}| }
 \hline
 \multicolumn{4}{|c|}{Results} \\
 \hline
 Measure& ref\_var\_ref\_close\_var&ref\_var\_ref\_distant\_var&mw\_size\\
 \hline
 diffint   & \textbf{6.231}   & 7.853&2\\
 citydist  & \textbf{8.836}  &   11.213&2\\
 corrdist  &\textbf{0.503}&   2.378&2\\
\hline
 diffint   & \textbf{2.163}    & 4.782&3\\
 citydist  & \textbf{4.556}  &   7.181&3\\
 corrdist  &\textbf{0.713}&   4.044&3\\

 \hline
\end{tabular}
\caption{\textit{Euclidean distance between similarity scores}} \label{tab:eudist}
\end{table}

Overall, the results of the motif embeddings show that vector representations of folksong motifs capture contextual melodic features. Query results  show how motifs can be modeled with the skip-gram version of the word2vec from monophonic contexts. One of the advantages of this method is that motifs can be easily modeled in a complete unsupervised manner given a context, and they can be retrieved using the cosine distance. At the same time, with large corpora the algorithm tends to discover multiple motifs, some of which may be irrelevant for the musicological analysis.

\section{Conclusions}

Word2vec has been used to model complex Western polyphonic classical music \cite{herremans2017modeling}. In this article the skip-gram version of word2vec is used to learn rich representations of monophonic motifs from the Essen folksong collection. The proposed approach shows how motifs from folksongs can be learned from a large corpus and compared with each other using the cosine similarity. This approach can be very useful for the musicological study of folksong variation using small melodic units such as motifs. It also shows, how word2vec is able to capture and model melodic contexts from monophonic songs. Future work should concentrate on the filtering of motifs based on different musicological criteria, to avoid a combinatorial explotion and to select relevant motifs for the musical analysis.

The evaluation of WE is an important research topic in the NLP literature \cite{schnabel2015evaluation}. In this article a novel computational method for evaluating the quality of motif embeddings is proposed. The approach presented shows how the model captures different degrees of motif similarity. This evaluation method can be very useful for studying the similarity of melodic segments based on motifs and their related variants. Future work in this area should include a cognitive similarity evaluation task performed by human participants to test the quality of the embeddings.

%

\bibliographystyle{splncs04}
\bibliography{mybibliography}
\end{document}